\documentclass{aastex63}
\usepackage{upgreek}

\newcommand\aastex{AAS\TeX}

\shorttitle{A compositional gradient in the protoplanetary disk?}
\shortauthors{Marsset et al.}

\graphicspath{{./}{figures/}}

\begin{document}

\title{Template \aastex Article with Examples: 
v6.3\footnote{Released on June, 10th, 2019}}

\correspondingauthor{Michael Marsset}
\email{mmarsset@eso.org}

\title{Col-OSSOS: Evidence for a compositional gradient inherited from the protoplanetary disk?}

\author[0000-0001-8617-2425]{Micha$\ddot{\rm e}$l Marsset}
\affiliation{European Southern Observatory, Alonso de C\'ordova 3107, Santiago, Chile}
\affiliation{Department of Earth, Atmospheric and Planetary Sciences, MIT, 77 Massachusetts Avenue, Cambridge, MA 02139, USA}

\author[0000-0001-6680-6558]{Wesley C. Fraser}
\affiliation{Herzberg Astronomy and Astrophysics Research Centre, National Research Council, 5071 W. Saanich Rd. Victoria, BC, V9E 2E7, Canada}
\affiliation{Department of Physics and Astronomy, University of Victoria, Elliott Building, 3800 Finnerty Road, Victoria, BC V8P 5C2, Canada}

\author[0000-0003-4365-1455]{Megan E. Schwamb}
\affiliation{Astrophysics Research Centre, Queen's University Belfast, Belfast BT7 1NN, United Kingdom}

\author[0000-0002-8032-4528]{Laura E. Buchanan}
\affiliation{Astrophysics Research Centre, Queen's University Belfast, Belfast BT7 1NN, United Kingdom}

\author[0000-0003-4797-5262]{Rosemary E. Pike}
\affiliation{Institute of Astronomy and Astrophysics, Academia Sinica; 11F of AS/NTU Astronomy-Mathematics Building, No.1, Sec. 4, Roosevelt Rd, Taipei 10617, Taiwan, R.O.C.}

\author[0000-0002-6830-476X]{Nuno Peixinho}
\affiliation{Instituto de Astrof\`{i}sica e Ci\^{e}ncias do Espa\c{c}o, Universidade de Coimbra, 3040-004 Coimbra, Portugal}

\author[0000-0001-8821-5927]{Susan Benecchi}
\affiliation{Planetary Science Institute, 1700 East Fort Lowell, Suite 106, Tucson, AZ 85719}

\author[0000-0003-3257-4490]{ Michele T. Bannister}
\affiliation{School of Physical and Chemical Sciences | Te Kura Mat\=u University of Canterbury, Private Bag 4800, Christchurch 8140, New Zealand}

\author[0000-0001-6541-8887]{Nicole J. Tan}
\affiliation{School of Physical and Chemical Sciences | Te Kura Mat\=u University of Canterbury, Private Bag 4800, Christchurch 8140, New Zealand}

\author[0000-0001-7032-5255]{J.J. Kavelaars}
\affiliation{Herzberg Astronomy and Astrophysics Research Centre, National Research Council, 5071 W. Saanich Rd. Victoria, BC, V9E 2E7, Canada}
\affiliation{Department of Physics and Astronomy, University of Victoria, Elliott Building, 3800 Finnerty Road, Victoria, BC V8P 5C2, Canada}

\begin{abstract}

In the present-day Kuiper Belt, the number of compositional classes and the orbital distributions of these classes hold important cosmogonic implications for the Solar System. 
In a companion paper by Fraser et al., we demonstrate that the observed color distribution of small (H$\gtrapprox$6) Trans-Neptunian Objects (TNOs) 
can be accounted for by the existence of only two composition classes, named \textit{brightIR} and \textit{faintIR},  
where the range of colors in each class is governed by a mixture of two material end members. 
Here, we investigate the orbital distribution of the two color classes identified by Fraser et al. and find that the orbital inclinations of the \textit{brightIR} class objects are correlated with their optical colors. 
Using the output of numerical simulations investigating the orbital evolution of TNOs during their scattering phase with Neptune, we show that this correlation could reflect a composition gradient in the early protoplanetary disk, in the range of heliocentric distances over which TNOs from the \textit{brightIR} class accreted. 
However, tensions between this interpretation and the existence of blue contaminants among cold classical TNOs, and possible alternative origins for the detected correlation, currently bear uncertainty on our proposed interpretation. 
\\

\end{abstract}


\section{Introduction} \label{sec:intro}

Trans-Neptunian Objects (TNOs), the population of small bodies located beyond 30 astronomical units (au) in the Kuiper belt, represent a relic of the original protoplanetary disk that formed around our Sun 4.6~Gyrs ago. 
The composition and orbital properties of these objects have long been a topic of interest to the scientific community, as they hold clues to the formation and past evolution of our Solar System. 
In particular, the number of compositional classes existing in the TNO population is believed to reflect the level of compositional stratification that occurred in the early Solar System, before the dispersal of the protoplanetesimal disk. 
In addition, it is now widely accepted that the present-day orbital distribution of this population was shaped very early by the migrations of the giant planets (e.g., \citealt{Malhotra:1995dy, Gomes:2003cd, Levison:2008, Morbidelli:2020, Pirani:2021}). 
Deciphering both the color and orbital distributions of TNOs is thus key to understanding the first million to hundred million years of Solar System evolution, and the mechanisms that led to the present-day configuration. 

This past decade, the number of compositional classes existing in the outer Solar System has been a matter of debate, with three \citep{Fraser:2012, Pike:2017, Schwamb:2019} to ten classes \citep{DalleOre:2013} being proposed. 
The Colours of the Outer Solar System Origins Survey (Col-OSSOS; \citealt{Schwamb:2019}), a large survey on the Gemini North and the Canada-France-Hawaii Telescope (CFHT) observatories, was designed, among other objectives, to settle this debate. 
Col-OSSOS acquires high-quality near-simultaneous visible and near-infrared photometric observations of a magnitude-limited sample of TNOs discovered by the Outer Solar System Origins Survey (OSSOS; \citealt{Bannister:2016, Bannister:2018}). 
After 6 years of operations and 92 observed TNOs, we report in a companion paper by Fraser et al. evidence for only two composition classes in the Kuiper belt, thereby pointing towards limited radial compositional stratification in the early outer Solar System.
In the toy model proposed by Fraser et al., TNOs from each of the two classes -- named \textit{brightIR} and \textit{faintIR} based on their reflectivity in the near-infrared -- span a wide range of optical and near-infrared colors that is governed by a mixture of two material end members, with the blue-side end member being shared by the two classes. 
Most color properties of the TNO population reported by previous studies, including the known optical bifurcation into the so-called \textit{less red} and \textit{very red} color groups (e.g., \citealt{Fraser:2012, Peixinho:2012, Tegler:2016, Wong:2017, Marsset:2019}), and the tendency for cold classicals -- the population of TNOs on near-circular and low-inclination i$<$5$^\circ$ orbits \citep{Tegler:2000} -- to occupy a distinct color space than their excited counterparts \citep{Pike:2017}, can be accounted for by this two compositional-class model.

Previous works have shown that outer Solar System colors correlate with orbital elements (e.g., \citealt{Tegler:2000, Peixinho:2008, Tegler:2016, Marsset:2019, AliDib:2021}), which was interpreted as evidence for the existence of multiple classes of objects with distinct origins \citep{Marsset:2019}. 
Motivated by these earlier results, we investigate here the existence of similar correlations in the new {\it faintIR/brightIR} taxonomic scheme, using a combined dataset of TNO color measurements from Col-OSSOS and previous observations (Section~\ref{sec:sample}). 
In addition to the previously known concordance between color classes and dynamical classes, we report the very first detection of an \textit{intraclass} color versus inclination correlation, in the group of \textit{brightIR} TNOs (Section~\ref{sec:correlations}). 
Using the output of numerical particle integrations that simulate the orbital evolution of TNOs during their scattering phase with Neptune, we show that this correlation can be reproduced assuming that TNO colors were set by their original formation location in the Kuiper belt (Section~\ref{sec:origins}). 
Alternative explanations for the origin of the color-inclination correlation, however, currently cannot be ruled out.

\section{TNO sample}
\label{sec:sample}

\subsection{Sample definition}
\label{sec:def}

Observing conditions of the Col-OSSOS dataset analysed in this work are provided in the companion paper of Fraser et al. 
In order to increase our sample size and strengthen any existing correlation in our dataset, we combine this dataset with additional, previously-published measurements of similar photometric quality acquired with the \textit{Hubble Space Telescope} (HST) as part of the Hubble/WFC3 Test of Surfaces in the Outer Solar System (H/WTSOSS; \citealt{Fraser:2012, Fraser:2015}). 

We consider both visible and near-infrared colors, i.e. (g-r) and (r-J) colors for the Col-OSSOS dataset, and (F606w–F814w) and (F814w-F139m) for the H/WTSOSS dataset. 
In total, this represents 198 photometric measurements of 189 outer Solar System objects collected over the full visible+near-infrared wavelength range, with a few objects having multiple measurements. 
One object from the HST sample, 2007~OR10, was rejected from our dataset 
because it is a dwarf planet (H=1.9) with a distinct (methane-rich) surface composition and rheology from small TNOs. 

To obtain an homogeneous dataset, all color measurements obtained in the different filters were converted to spectral slope (\textit{s}) using the Synphot tool in the STSDAS software package\footnote{www.stsci.edu/institute/software\_hardware/stsdas}, following the method of \citet{Fraser:2017}. 
In what follows, \textit{s} is expressed in terms of percentage increase in reflectance per thousand angstr\"oms  change in wavelength (\%/$10^3\,\rm{\AA}$), normalized to unity at 550\,nm. 
A slope average was obtained for each object observed multiple times, weighting by the invert squared uncertainty on the individual slope measurements. 
The physical and orbital properties and spectral slopes measurements of the complete set of TNOs are provided in Appendix\,\ref{sec:appA}. 

\subsection{Taxonomic classification}
\label{sec:taxonomy}

Objects were then classified into the \textit{brightIR} and \textit{faintIR} classes using the color projection system of Fraser et al., i.e., using their location distance on the parallel ($PC^1$) and perpendicular ($PC^2$) vectors along the so-called \textit{reddening curve}. 
The reddening curve is the curve of equal spectral slope in all colors, meaning that, in a given color-color plot, an object located below the reddening curve has a concave (frown) spectrum, whereas an object located above is spectrally convex (smile).
Following Fraser et al., Col-OSSOS objects with $PC^2_{grJ}>-0.13$, and those with $PC^2_{grJ}<-0.13$ and $PC^1_{grJ}<0.40$ were classified as \textit{brightIR}, whereas objects with $PC^2_{grJ}<-0.13$ and $PC^1_{grJ}>0.40$ were classified as \textit{faintIR}.
HST objects were classified following the same criteria, using data from photometric filters F606w, F814W and F139m instead of g, r and J, respectively. 
A few objects in our dataset were visited several times and have multiple optical and NIR color measurements collected over the different observing epochs. Two of them, 2006~QP180 and 2013~JK64 have inconsistent surface classification between the visits (see discussion in companion paper by Fraser et al.). We choose not to attribute any class to these objects and do not include them in our analysis. \textcolor{black}{However, we verified that putting these objects either in the \textit{brightIR} or the \textit{faintIR} class does not change the results of our statistical analysis in any way.}

\section{Correlations in the class of brightIR TNOs}
\label{sec:correlations}

\subsection{Color versus inclination distributions}
\label{sec:col_inc}

\subsubsection{Col-OSSOS+H/WTSOSS}
\label{sec:colhst_corr}

\begin{figure}[h!]
\centering
\includegraphics[angle=0, width=0.45\linewidth, trim=0cm 0cm 0cm 0cm, clip]{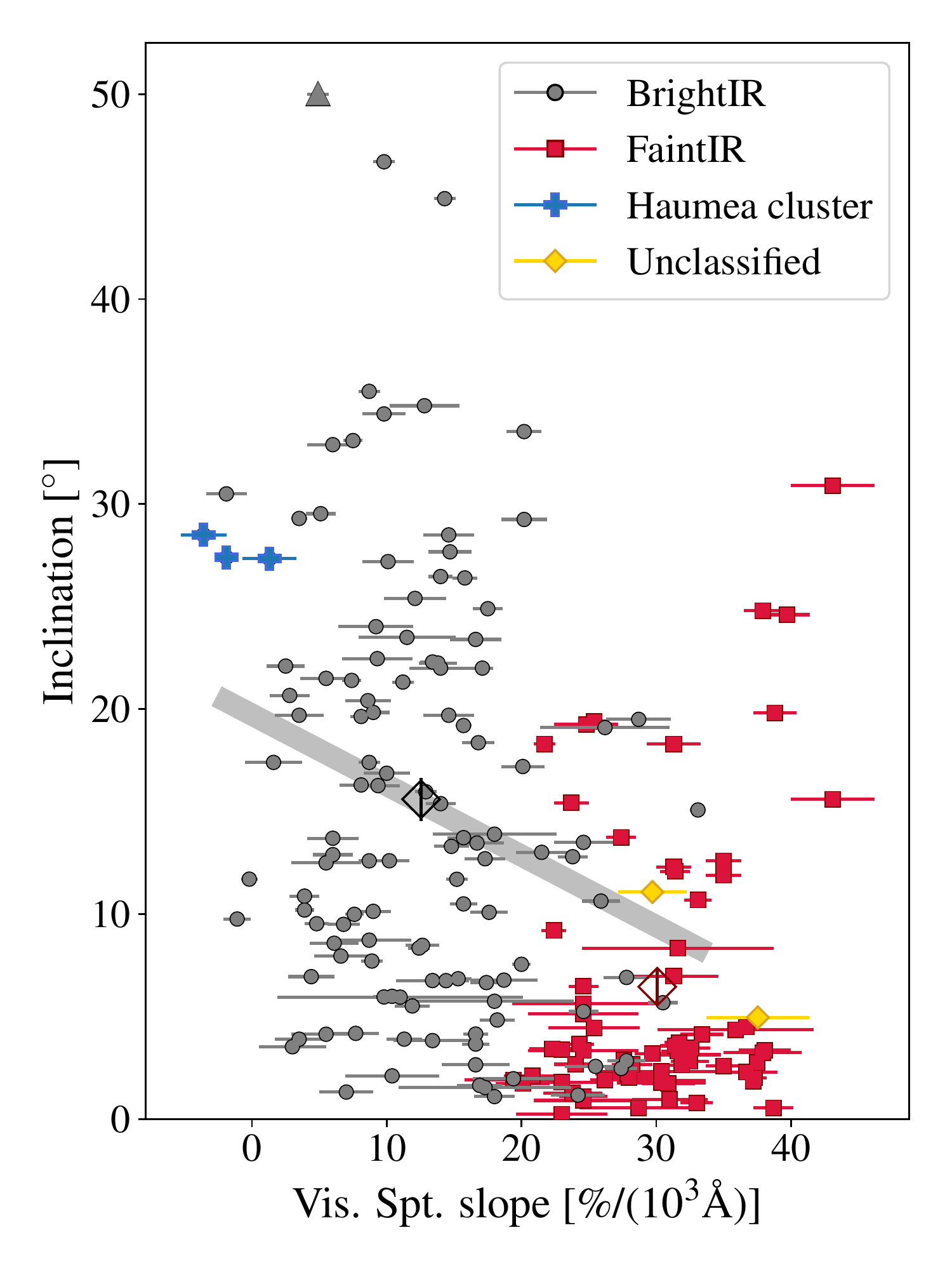}
\includegraphics[angle=0, width=0.45\linewidth, trim=0cm 0cm 0cm 0cm, clip]{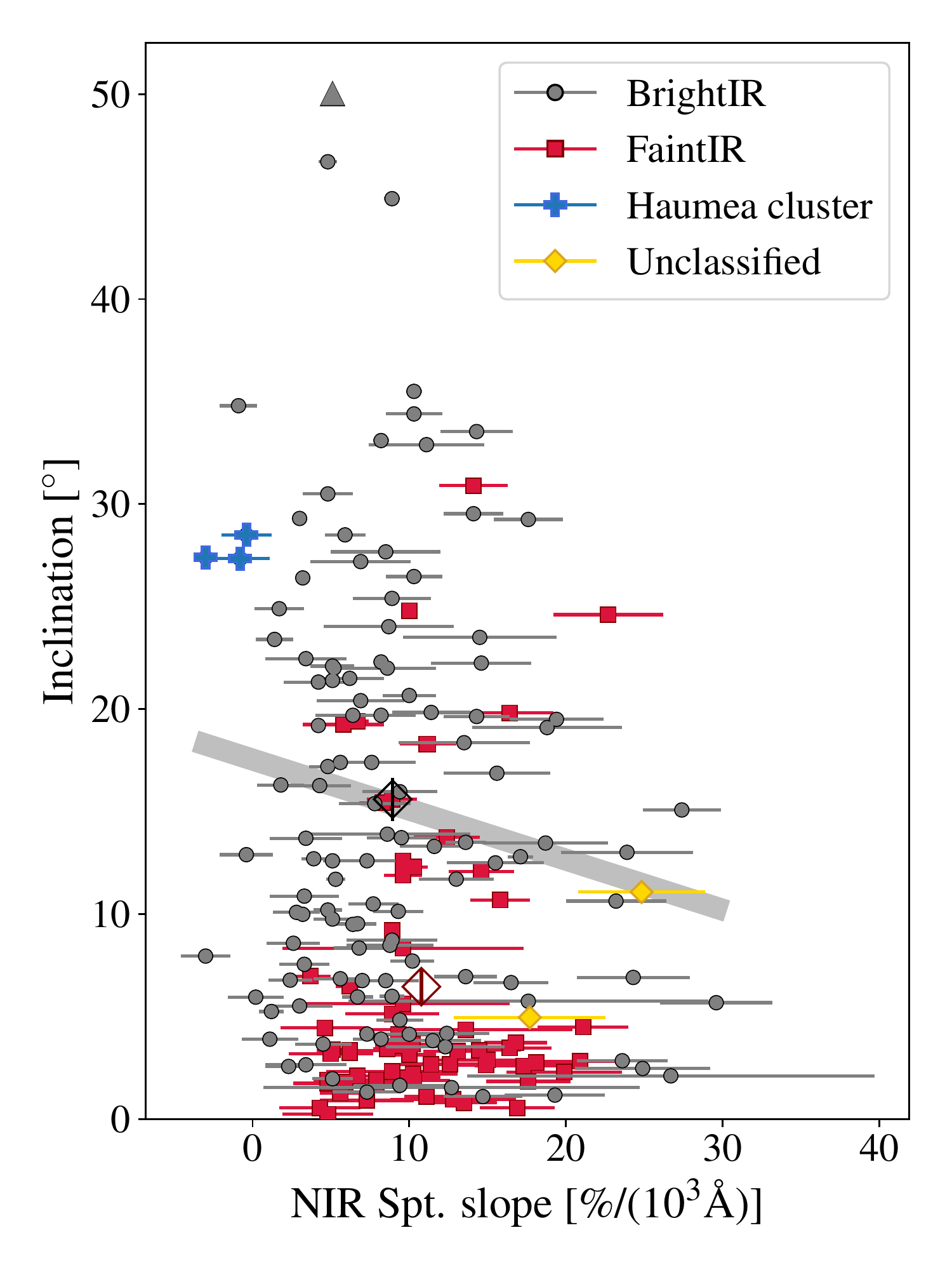}
 \caption{Orbital inclination ($i$) versus visible (left) and near-infrared (right) spectral slope (\textit{s}) of the \textit{brightIR} (grey circles) and \textit{faintIR} (red squares) classes of TNOs in the Col-OSSOS+H/WTSOSS dataset. Two objects in our dataset (golden diamonds) were visited twice and have inconsistent surface classification between the two visits. We choose not to attribute any class to these objects and do not include them in our analysis. (127546) 2002~XU93, with $i$=77.89$\degr$, is plotted as a upward grey triangle at $s_{vis}$=4.9\%/$10^3\,\rm{\AA}$, $s_{nir}$=5.1\%/$10^3\,\rm{\AA}$, inc=50$\degr$ for better readability. 
 The black and red empty diamonds indicate the average inclination values of the \textit{brightIR} and \textit{faintIR} classes, respectively, with y-axis errorbars corresponding to the uncertainty on the average. 
 \textit{BrightIR} and \textit{faintIR} TNOs exhibit distinct inclination distributions, as previously found by \cite{Marsset:2019} in the case of \textit{less red} and \textit{very red} TNOs. 
 The grey lines are linear fits to the \textit{s} versus $i$ distributions of the \textit{brightIR} class, with the exclusion of bodies located in the orbital space of the Haumea cluster (blue crosses). The negative slope suggests an anti-correlation between the two variables, both in the visible and near-infrared wavelength ranges. The statistical robustness of the correlations is discussed in Section~\ref{sec:colhst_corr}.}
\label{fig:slope_vs_i}
\end{figure}

In Fig.\,\ref{fig:slope_vs_i}, we show the orbital inclination versus optical and near-infrared color slope of our dataset. 
Similarly to \cite{Marsset:2019}, who showed the existence of distinct inclination distributions for the \textit{less red} and \textit{very red} TNO populations, Fig.\,\ref{fig:slope_vs_i} highlights the distinct orbital inclinations for the \textit{brightIR} and \textit{faintIR} classes. 
The fact that the same trend is observed in both classification systems is not surprising: both the \textit{brightIR} and less red classes and the \textit{FaintIR} and very red classes largely overlap in object membership. If placing the optical slope limit between \textit{less red} and \textit{very red} objects at 20.6\%/$10^3\,\rm{\AA}$ as in \citet{Marsset:2019}, 87\% of \textit{brightIR} measurements in our dataset (111 out of 127 measurements) would fall in the \textit{less red} category, while 96\% of the \textit{faintIR} ones (68 out of 71) would fall in the \textit{very red} category. 
Our dataset also shows distinct eccentricity distributions of the two composition classes, which is mostly due to the presence of dynamically-quiescent cold classical TNOs in the \textit{faintIR} class. 

We then consider each class separately and search for additional intraclass correlations between spectral slopes and orbital parameters by means of the Pearson and Spearman $r$ tests. 
The likelihood of the correlations was evaluated by bootstrapping with replacement $10^5$ times values of spectral slope and orbital parameter taken from the observed classes, independently. 
To take our photometric accuracy into consideration, each spectral slope value of the bootstrapped populations was scattered by a probabilistic value taken from a Gaussian distribution with width taken from a random color uncertainty in our sample. 
The Pearson and Spearman correlation values were then recorded for each of the $10^5$ simulated populations. 
The results of the statistical tests are provided in Appendix~\ref{sec:appB}.

Notably, we find that only 0.5\% of the simulated populations bootstrapped from the \textit{brightIR} class exhibit a stronger visible slope versus orbital inclination correlation than the real (observed) \textit{brightIR} class, implying a confidence level (CL) of 99.5\% that the observed negative correlation is not due to chance. 
When rejecting one extremely-high inclination TNO from the \textit{brightIR} class, (127546) 2002~XU93 with i=77.89$\degr$, that could have potentially biased the sample, we find that the correlation remains significant, with CL=99.1\%.
The statistical significance of the correlation, however, weakens (CL=97.9\%), when rejecting objects located in the orbital region of the Haumea collisional family (hereafter referred to as ``HCs" for ``Haumea Candidates"), between semi-major axes $40.5<a<46$~au and inclinations $25<i<29^{\circ}$. 
Haumea family members are located on high-inclination orbits and are known to exhibit distinct, ice-rich surfaces compared to the bulk population of TNOs (e.g., \citealt{Schaller:2008, Snodgrass:2010}). As such, including these objects in our dataset automatically enhances the colour versus inclination correlation.

Due to the precession of TNO orbits around their local forcing pole, the osculating ecliptic inclination ($i$) of these objects varies over time. This is why the conserved quantity of ``free inclination" ($i_{free}$) was introduced and is often more relevant when studying TNO orbits than the ecliptic inclination (e.g., \citealt{Brown:2004,Elliot:2005, Chiang:2008,Volk:2017,vanLaerhoven:2019,Huang:2022}). 
We investigated the visible slope versus inclination anti-correlation detected in our dataset using free inclination values from \citet{Huang:2022}. Specifically, if an object in our dataset had its $i_{free}$ value computed by these authors, we used that value when performing our statistical test, otherwise we used $i$. By doing so, we find a similar CL of 99.3\% (versus CL=99.5\% for the osculating element) that the visible slope versus orbital inclination anti-correlation of the \textit{brightIR} objects is not due to chance, and 97.2\% (versus 97.9\% in the osculating case) when HCs are excluded from the dataset. In short, considering $i_{free}$ instead of $i$ does not reveal any new correlation, nor strengthen the existing ones in our dataset. Therefore, in what follows we use values of osculating inclination in our statistical analysis, unless explicitly stated otherwise.

In the new classification system of Fraser et al., optical and near-infrared colors are correlated in each compositional class. 
A similar correlation between colors and inclinations to the one observed in the visible should therefore also be seen in the near-infrared. 
In Fig.\,\ref{fig:slope_vs_i}, we observe such a trend of decreasing orbital inclination as a function of increasing near-infrared spectral slope values, but the trend appears weaker than in visible wavelengths, which is confirmed when applying the Pearson and Spearman correlation tests (CL=94.1 and 81.9\%, respectively, and lower when HCs are rejected). 
The lack of a statistically strong correlation in the near-infrared may relate, in part, to the larger uncertainties on the near-infrared photometric measurements (average uncertainty 2.4\%/$10^3\,\rm{\AA}$) 
compared to visible ones (1.9\%/$10^3\,\rm{\AA}$). 

Finally, we note a weak positive correlation between visible slope and orbital inclination in the \textit{faintIR} class (CL=\textcolor{black}{98.4/92.8 }\% for Pearson/Spearman). 
We also note a strong correlation between the visible slopes and eccentricities of the \textit{faintIR} class in the H/WTSOSS-only dataset (CL=\textcolor{black}{99.3}\%). 
However, these correlations are driven by the inclusion of Centaurs in the datasets, objects whose current orbital elements (especially the eccentricity) are affected by recent scattering events with Neptune, and whose surface colors are preferentially altered compared to TNOs as their orbits evolve closer to the Sun. 
Once Centaurs are removed, the correlations become statistically non significant. 
Therefore, we do not consider these correlations in the \textit{faintIR} class to be meaningful. 

\subsubsection{Minor Bodies in the Outer Solar System (MBOSS)}
\label{sec:mboss_corr}

We further verified whether a similar correlation between spectral slope and inclination could be found in previous TNO datasets when placed into the {\it faintIR/brightIR} classification scheme. 
Unfortunately, few TNOs had their photometric properties measured over both the visible and NIR wavelengths before Col-OSSOS and H/WTSOSS. 
The only substantial dataset available is the {\it Minor Bodies in the Outer Solar System} (MBOSS) dataset \citep{Hainaut:2002,Hainaut:2012}, but visible and NIR measurements from that survey were not acquired simultaneously, making taxonomic classification uncertain. 
Nevertheless, we classified objects into the {\it brightIR} and {\it faintIR} classes based on their $PC^1_{VRJ}$ and $PC^2_{VRJ}$ values using the same approach as previously described for the Col-OSSOS+H/WTSOSS dataset, and converted their (V-R) and (R-J) colors into visible and NIR spectral slope values. 
Dwarf planets with $H_{mag}<$3 were rejected from the sample. 

\begin{figure}[h!]
\centering
\includegraphics[angle=0, width=0.45\linewidth, trim=0cm 0cm 0cm 0cm, clip]{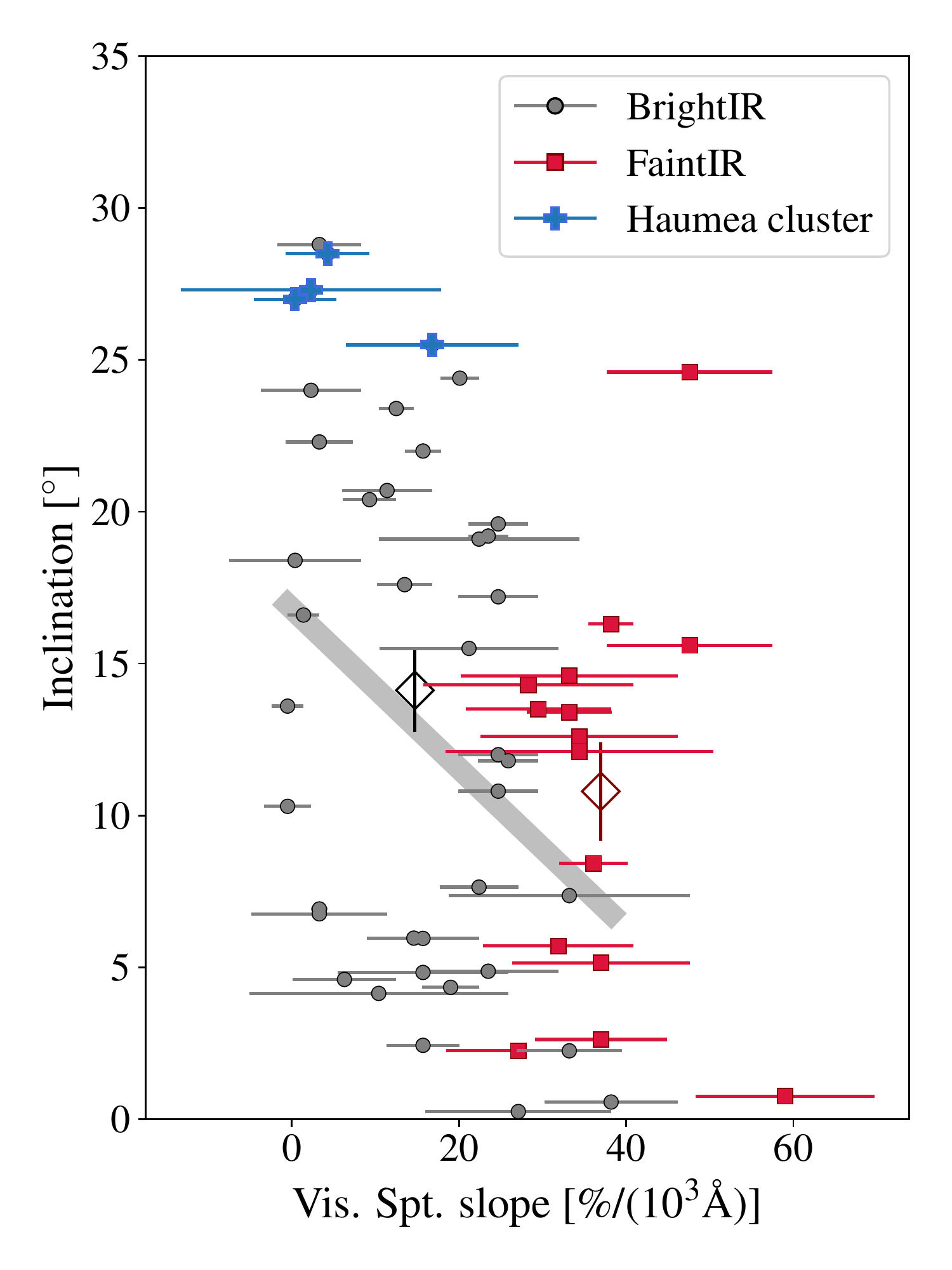}
\includegraphics[angle=0, width=0.45\linewidth, trim=0cm 0cm 0cm 0cm, clip]{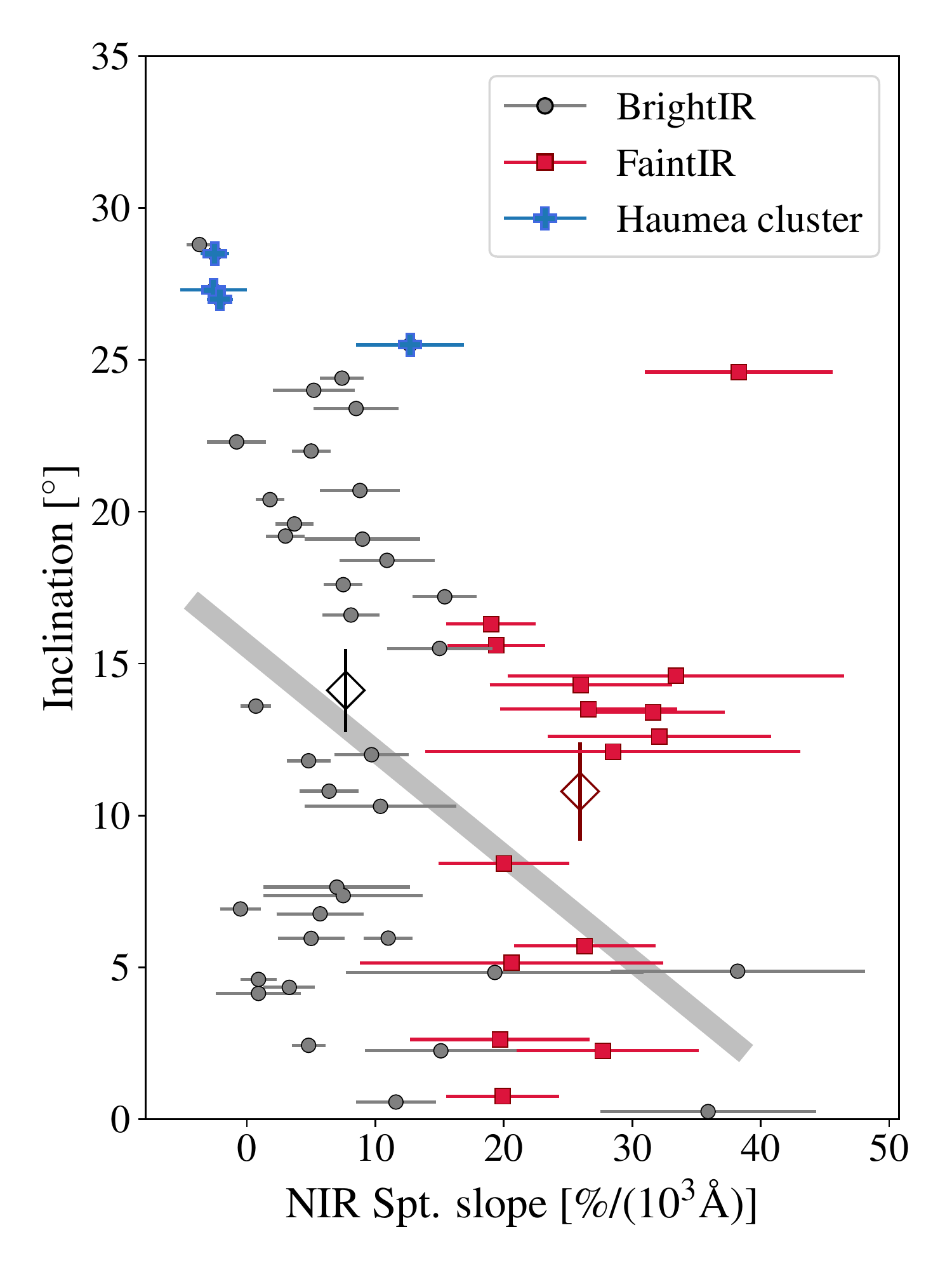}
 \caption{Same as Fig.\ref{fig:slope_vs_i} for (V-R) and (R-J) colour measurements from the MBOSS dataset converted to spectral slopes. 
 As in the Col-OSSOS+H/WTSOSS dataset, \textit{brightIR} objects from MBOSS exhibit a trend of decreasing inclination as a function of increasing visible (left) and NIR (right) spectral slopes. However, the significance of the correlation is lower than for the Col-OSSOS+H/WTSOSS dataset, see Section\,\ref{sec:mboss_corr}.} 
\label{fig:slope_vs_i_MBOSS}
\end{figure}

The resulting classified dataset is plotted in spectral slope versus orbital inclination in Fig.~\ref{fig:slope_vs_i_MBOSS}. 
{\it BrightIR} objects from this dataset exhibit the same anti-correlation between spectral slopes and inclinations as the Col-OSSOS+HST dataset, both in visible and NIR wavelengths. 
The probability that these correlations are due to chance, however, is high (\textcolor{black}{0.8}\% and \textcolor{black}{4.8}\% in the visible and NIR, respectively). When removing HCs, the correlations are statistically not significant. MBOSS also includes several large ($H_{mag}<$5) TNOs that may exhibit distinct (volatile-rich) surfaces from the bulk population of smaller TNOs. If removed from the dataset, no correlation remains (Appendix\,\ref{sec:appB}). 

\subsubsection{Col-OSSOS+H/WTSOSS+MBOSS}

Next, in order to investigate further the observed trends and correlations, we combined the three datasets (Col-OSSOS, H/WTSOSS and MBOSS) together and repeated our statistical tests using the combined dataset (Appendix\,\ref{sec:appB}). 
In this dataset, the CL for a correlation between spectral slope and orbital inclination is close to 100\% and 99.4/95.9\% (Pearson/Spearman) in the visible and NIR wavelengths, respectively. This correlation remains highly significant in the visible when removing HCs only (CL=99.9\%), and when removing HCs, large TNOs ($H_{mag}<$5) and Centaurs (CL=99.4/99.8\% for Pearson/Spearman).

We further tested the detected correlations against possible biases coming from additional dynamical and/or compositional groups in our dataset, as well as possible selection effects. 
First, we tested the statistical robustness of the correlation against possible class assignment errors. In the {\it brightIR/faintIR} scheme, low-$PC^1$ objects (i.e., those located at the blue-visible/blue-NIR corner of the color-color plot) are taxonomically the most ambiguous. This is because the color bifurcation of TNOs into the \textit{brightIR} and \textit{faintIR} classes occurs near $PC1>0.4$, or {\it (g-r})$>$0.8. Below this value, the two classes merge in {\it grJ} colors, which is why Fraser et al. suggested that these classes share the same blue material end member. Here, following Fraser et al., we chose to classify objects with $PC^2<-0.13$ and $PC^1<0.40$ as \textit{brightIR}, but some of these objects could belong to the \textit{faintIR} class. In order to test the robustness of the color-inclination correlations of the \textit{brightIR} class against that possibility, we repeated our statistical tests after rejecting objects with $PC^2<-0.13$ and $PC^1<0.40$ from that class. By doing this, we find that the visible slope versus orbital inclination correlation of the \textit{brightIR} class remains statistically significant, with only 0.2\% probability (1.8\% if excluding HCs) of being due to chance. Adding objects with $PC^2<-0.13$ and $PC^1<0.40$ to the \textit{faintIR} group does not reveal any new correlation in that group.

Traditionally, cold classical TNOs have been regarded as a distinct class of objects having a distinct origin from the rest of the TNO population, owing to their unique orbital and surface properties (e.g., \citealt{Tegler:2000, Brown:2001, Parker:2010, Fraser:2012, Pike:2017}), as well as their high fraction of binaries \citep{Noll:2008}. 
As a consequence, it has been a common practice to remove any object located on a cold classical orbit from studies of the orbital and compositional distributions of the \textit{less red} and \textit{very red} classes of TNOs, to avoid any cold-classical contaminant in these classes (e.g., \citealt{Marsset:2019}). 
In the new {\it faintIR/brightIR} classification scheme, however, cold classicals are merely the dynamically-quiescent (low-$i$ and low-$e$) end tail of the {\it faintIR} TNO class. 
Therefore, it appears inappropriate to reject low-$i$ objects when studying the \textit{brightIR} and \textit{faintIR} classes. 
However, for historical consistency, let's assume that low-$i$ objects still constitute their own class of objects. 
We rejected them from our dataset using \citet{Huang:2022}'s free inclination values and the following orbital cuts for cold classicals: semi-major axes $42.5<a<45$~au and $i_{free}<4^{\circ}$ or $45<a<47$~au with $i_{free}<6^{\circ}$, because the outer part of the classical belt is thought to have experienced slightly more dynamical stirring. 
When doing so, we find that the Pearson and Spearman $r$ correlation test return a CL of \textcolor{black}{99.4}\% (96.4\% and 94.7\% without HCs, respectively) that the inclination \textit{versus} spectral slope correlation of the \textit{brightIR} objects still exists. 

We further note that the detected correlation is statistically stronger when dividing our sample following the {\it faintIR/brightIR} classification scheme, than when dividing it into the \textit{less red} and \textit{very red} categories as performed in previous studies (e.g., \citealt{Peixinho:2012, Tegler:2016, Wong:2017, Marsset:2019}). Using the classic classification method into \textit{less red} and \textit{very red} objects with a slope transition value of $20.6\%10^3\,\rm{\AA}$ \citep{Marsset:2019}, the CL of visible slopes being correlated to orbital inclinations in our dataset is \textcolor{black}{92.4}\% for the \textit{less red} category without HCs, i.e., significantly lower than the correlation detected using the {\it faintIR/brightIR} classification scheme (CL=99.9\% for the combined dataset without HCs). 
In other words, the {\it faintIR/brightIR} classification scheme reveals new color versus dynamical structure in the Kuiper belt that could not be found by earlier studies. 
The main disadvantage of this classification system relies on the requirement of observations in both the optical and near-infrared wavelength ranges, the later being rather expensive in terms of telescope time. 
No additional correlation could be found between the visible and near-infrared spectral slopes of the \textit{brightIR} and \textit{faintIR} classes and their orbital elements 
in the combined dataset (Appendix\,\ref{sec:appB}). 

\subsection{\textcolor{black}{$PC$ versus inclination distributions}}
\label{sec:pc_inc}

\begin{figure}[h!]
\centering
\includegraphics[angle=0, width=0.45\linewidth, trim=0cm 0cm 0cm 0cm, clip]{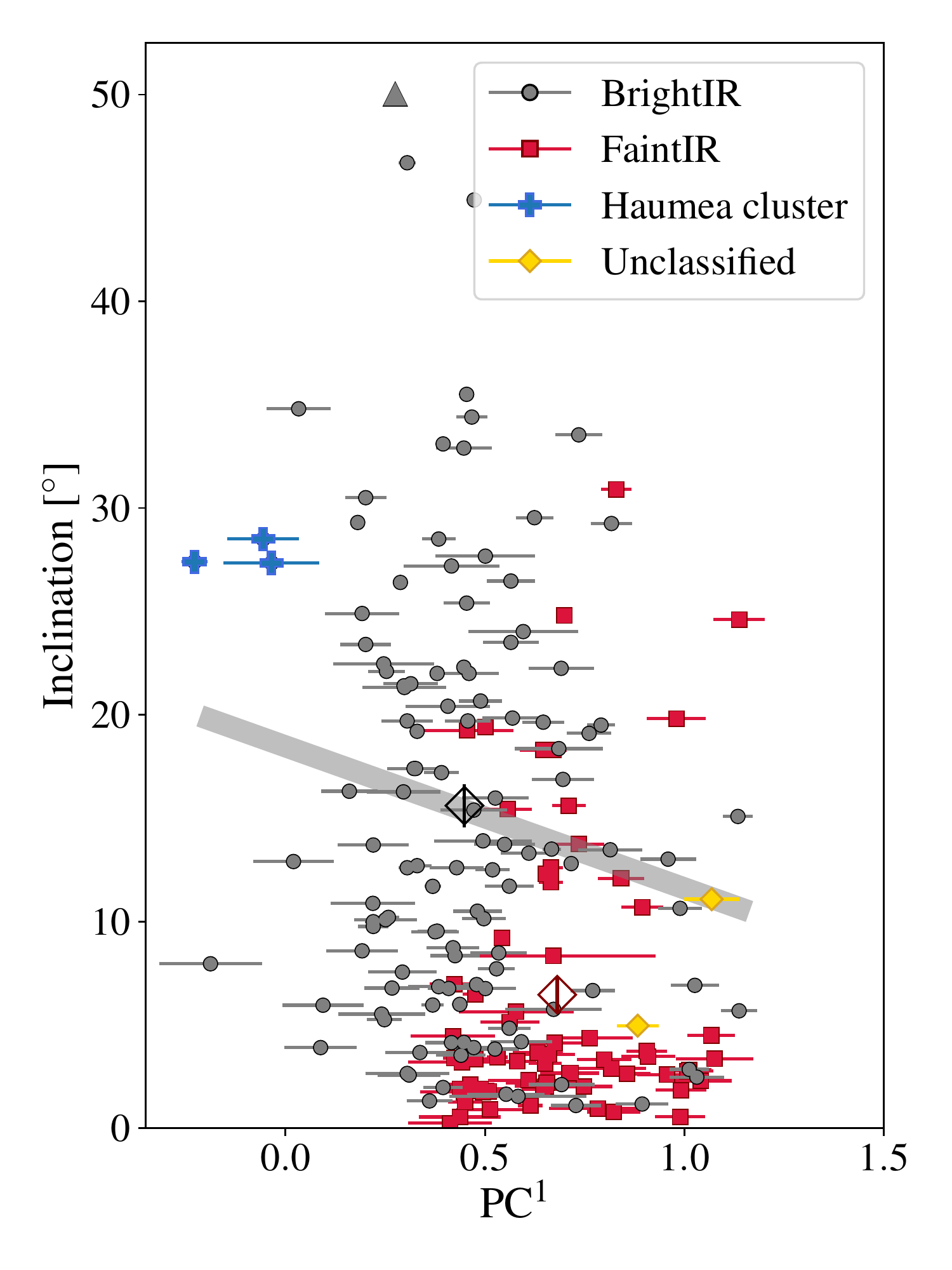}
\includegraphics[angle=0, width=0.45\linewidth, trim=0cm 0cm 0cm 0cm, clip]{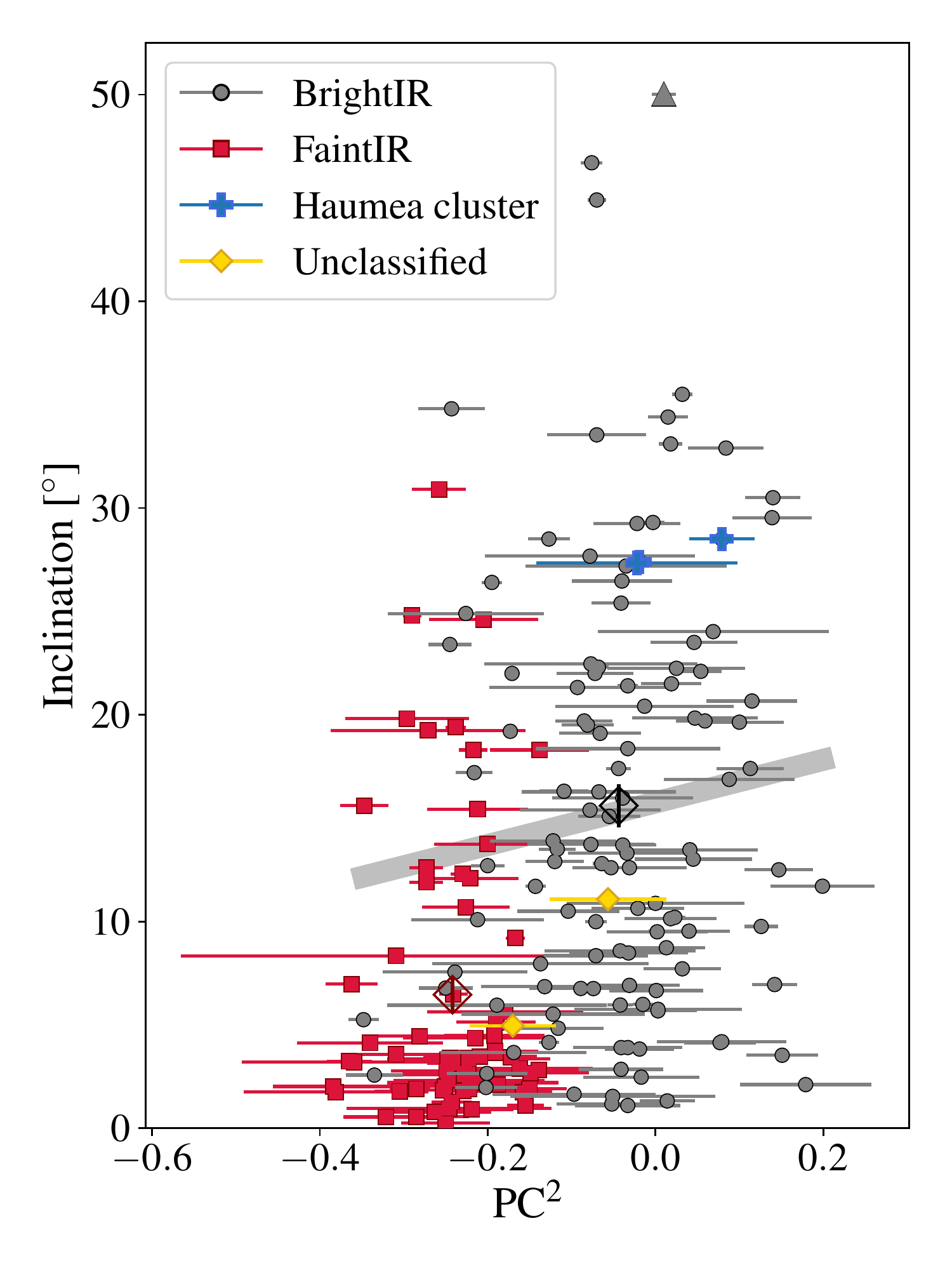}
 \caption{\textcolor{black}{Orbital inclination ($i$) versus PC values (i.e., location distance on the parallel ($PC^1$) and perpendicular ($PC^2$) vectors along the reddening curve) of the {\it brightIR} and {\it faintIR} classes of TNOs in the Col-OSSOS+H/WTSOSS dataset. The color and symbol code is as in Fig.\,\ref{fig:slope_vs_i}. 
 In the {\it brightIR/faintIR} taxonomy, intraclass color variations (and $PC^1$ variations) are caused by varying fractions of the two material end-members composing the class. As such, the negative slope of the fitted grey line in the left panel ($PC^1$ versus inclination) indicates that {\it brightIR} objects on high-$i$ orbits tend to be made of larger fractions of the blue material end member compared to low-$i$ objects from that same class (and, symmetrically, low-$i$ {\it brightIR} objects contain larger fractions of the red material end member compared to high-$i$ {\it brightIR} objects). The positive slope of the dashed line in the right panel ($PC^2$ versus inc) reflects the fact that {\it brightIR} objects on low-$i$ orbits generally tend to have more concave spectra, whereas those on high-$i$ orbits are on average spectrally more convex. The statistical robustness of these correlations is discussed in Section~\ref{sec:pc_inc}.}}
\label{fig:pc_vs_i}
\end{figure}

\textcolor{black}{In the {\it brightIR/faintIR} taxonomy, intraclass color variations are caused by varying fractions of the two material end-members composing that class. 
The parameter $PC^1$, the parallel vector along the reddening curve, is a measurement of the abundance ratio of the two materials in that class. 
$PC^2$, the perpendicular vector along the reddening curve, is a measurement of the deviation from the reddening curve, i.e., a measurement of the concavity (negative $PC^2$) or convexity (positive $PC^2$) of the TNO spectra. 
Considering that: (1) $PC^1$ positively correlates with visible and NIR spectral slope, and (2) visible spectral slope negatively correlates with orbital inclination (Section~\ref{sec:col_inc}), a negative correlation between $PC^1$ and inclination is expected to be found in our dataset. 
In Fig.~\ref{fig:pc_vs_i}, we plot the Col-OSSOS+H/WTSOSS dataset in the $PC^1$ and $PC^2$ versus inclination parameter spaces. 
The negative slope of the linear fit to the distribution of {\it brightIR} objects supports the existence of the expected negative correlation between $PC^1$ and inclination in that class. 
Applying the Pearson and Spearman $r$ correlation tests to the combined Col-OSSOS+H/WTSOSS+MBOSS datasets returns a weak CL of \textcolor{black}{96.7}\% that this correlation is not due to chance. The CL becomes not significant when removing HCs from the dataset. 
If objects with $PC^2<-0.13$ and $PC^1<0.40$ are excluded from the class, the significance of the correlation increases to \textcolor{black}{99.8}\% (96.0\% without HCs). 
The weaker statistical significance of the correlation between $PC^1$ and inclination compared to that of the correlation between visible slope and inclination may be due to the fact that $PC^1$ values depend both on visible and NIR slope values. Yet, NIR slope measurements carry larger uncertainties compared to the visible ones, and they exhibit only a very weak trend with inclination. }

\textcolor{black}{Fig.~\ref{fig:pc_vs_i} further shows that $PC^2$ values globally increase as inclination values increase in the {\it brightIR} class, meaning that {\it brightIR} objects on low-inclination orbits generally tend to have more concave spectra, whereas those on high-inclination orbits are on average spectrally more convex. 
However, this correlation is statistically not significant: there is a \textcolor{black}{8.0}\% chance (9.7\% if rejecting HCs) of producing a stronger correlation by bootstrapping with replacement random distributions of objects from the observed distribution.} 

\section{Primordial origin?}
\label{sec:origins}

The wider inclination and eccentricity distributions of the \textit{brightIR} class compared to the \textit{faintIR} suggest that the former formed closer to the Sun and was subsequently dynamically more excited by Neptune's formation and migration \citep{Marsset:2019, Nesvorny:2020, AliDib:2021, Pirani:2021}. 
Similarly, the observed visible spectral slope versus inclination correlation measured in the \textit{brightIR} class, \textcolor{black}{and the weak $PC^1$ versus inclination trend in that same class,} can easily be understood if we assume that the current inclinations of the \textit{brightIR} TNOs primarily resulted from their degree of interaction with Neptune during the migration phase. 
In the scenario of a late phase of Neptune's outward migration, this degree of interaction is proportional to the formation distance of TNOs (e.g., \citealt{Nesvorny:2020, AliDib:2021}). 
This implies that the present inclinations and eccentricities of TNOs are expected to be linked to their initial semi-major axis, although additional dynamical processes, such as collisions and dynamical pushes during close encounters may have further altered their orbits. In the case of resonant objects, Kozai oscillations may also modify orbital elements \citep{Gomes:2005}.

\begin{figure}[h!]
\centering
\includegraphics[angle=0, width=1\linewidth, trim=0cm 0cm 0cm 0cm, clip]{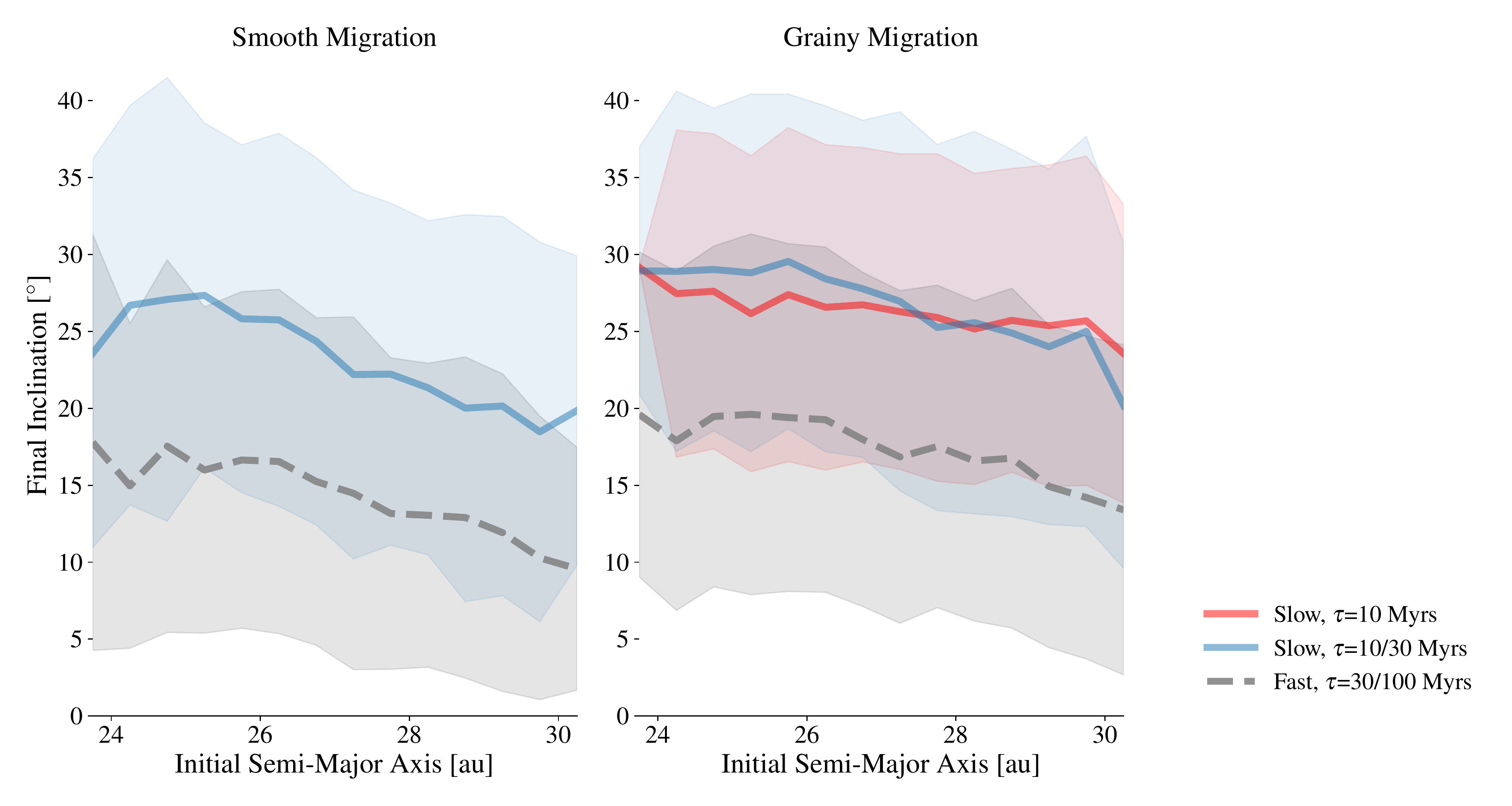}
 \caption{Final inclination distribution of test particles in the dynamical simulations of \citet{Kaib:2016} (blue and black lines) and \citet{Nesvorny:2016} (red line), as function of their initial semi-major axis (running average with 0.5-au steps). The left and right panels correspond to smooth and grainy migration scenarios, respectively. The continuous and dashed lines represent scenarios of fast and slow migrations, respectively, with exponential e-folding timescales ($\tau$) as indicated in the legend. When two $\tau$ values are provided, the first one corresponds to the value before Neptune's jump in semi-major axis, and the second one to the value after the jump. 
 Coloured regions correspond to the 1-$\sigma$ dispersion of inclination values at the final stage of the simulations. Despite the very large scatter in inclination values, the same trend of decreasing final inclinations as a function of increasing initial semi-major axis is observed in every simulations. \textcolor{black}{Combining all datasets, test particles have an average 1.0-degree decrease in final inclinations per 1-au increase in initial semi-major axis.}}
\label{fig:ai_vs_if}
\end{figure}

This is illustrated in Fig.~\ref{fig:ai_vs_if}, where we show the correlation between the initial semi-major axis and final inclinations of test particles in the dynamical models of \citet{Kaib:2016} (recently reanalysed by \citealt{Lawler:2019}) and \citet{Nesvorny:2016}. 
These models explore a range of scenarios for Neptune's migration, with different migration speeds and ``granularity" (small jumps) to simulate the presence of a large population (of a few thousands objects) of Pluto-sized ($\sim$1,000\,km in diameter) planetesimals. 
These models also include a large jump in semi-major axis and eccentricity in Neptune's migration, meant to simulate the scattering of a now-ejected ice giant \citep{Nesvorny:2015}. 
In the dynamical models of \citet{Kaib:2016}, Neptune's migration can either be slow, with exponential e-folding timescales of 10 and 30\,Myr before and after Neptune’s large jump, respectively, or fast, with e-folding timescales of 30 and 100\,Myr before and after the jump, respectively. 
In the simulation of \citet{Nesvorny:2016}, the migration is grainy and fast, with e-folding timescales of $\gtrapprox$10\,Myr. 
In each of these scenarios, the output population exhibits a trend of decreasing final inclinations as a function of increasing initial semi-major axis (Fig.~\ref{fig:ai_vs_if}), despite a large scatter in the inclination values.


Taken together, our observations and the dynamical simulations 
suggest the existence of a colour \textit{versus} formation distance gradient in the early proto-planetary disk, in the range of heliocentric distances over which the \textit{brightIR} TNOs originally formed. 
Taken in the context of the mixing model of Fraser et al., where intraclass color variations arise from the relative abundances of the two material end members that compose the class, this \textit{color}-distance gradient implies a \textit{composition}-distance gradient in the disk. 
Higher (or redder) spectral slopes in the outer Solar System are 
generally associated with higher content of organic matter so it follows that \textit{brightIR} objects formed at larger distances from the Sun would have earned a thicker irradiation crust. 
In addition, optically reddest \textit{brightIR} TNOs are the least common of the class, implying that either the density of surviving TNOs from the class drops off with formation distance, or that the red surfaces can only form for the most distant \textit{brightIR} objects (located closer to the \textit{brightIR}/\textit{faintIR} transition line).

Fig.~\ref{fig:ai_vs_if} further suggests that an object's orbital inclination is, on average, pumped up 1$^{\circ}$ more by Neptune scattering as its formation distance to the Sun is decreased by 1\,au. 
In our observed dataset (Fig.\,\ref{fig:slope_vs_i}), the linear fit to the color-inclination distribution spans 12 degrees for 95\% of the \textit{brightIR} TNOs, suggesting 
that the formation region of the \textit{brightIR} TNOs originally spanned across $\sim$12\,au in semi-major axis in the proto-planetary disk. 
Taking these numbers at face value would be unreasonable considering model and observation uncertainties. 
Nevertheless, the order of magnitude is entirely consistent with expectations for the original extension of the proto-planetary disk \citep{Nesvorny:2015b}, thereby providing additional credit to the hypothesis of a compositional gradient origin for the color-inclination distribution of the \textit{brightIR} objects.

Let us, however, point out some current tensions regarding this interpretation. 
First, it has been shown that the so-called blue binaries, a population of optically blue contaminants found in the orbital region of the otherwise optically red cold classicals, could only have moved a few astronomical units (au)  during the phase of Neptune's migration in order to preserve their 100\% binary fraction \citep{Fraser:2017, Fraser:2021,Nesvorny:2022}. In our proposed scenario, where the color versus inclination correlation stems from a composition gradient in the proto-planetary disk, this may pose a contradiction for the very blue optical and near-infrared colours (i.e., small $PC^1$ values) of the blue binaries, that would suggest a formation of these binaries at the inner edge of the proto-planetary disk, i.e. several au away from the transition line between the \textit{brightIR} and \textit{faintIR} populations \citep{Nesvorny:2020, AliDib:2021, Buchanan:2022}. Reconciling the presence of blue binaries among cold classical objects with our proposed idea of a composition gradient in the \textit{brightIR} formation region would require a precise measurement of the location of the transition line between the formation regions of \textit{brightIR} and \textit{faintIR} TNOs, and dynamical simulations of Neptune's migration able to displace the blue binaries across the full \textit{brightIR} formation region while preserving their binarity, low inclinations and low eccentricities. 

A second issue comes from the fact that our proposed scenario currently cannot be distinguished from another alternative: a collisional origin for the color versus inclination correlation. 
Impact-induced resurfacing could very well play a role because collision velocities -- and therefore resurfacing efficiency -- are a function of the mutual inclinations and eccentricities of the colliding bodies (e.g., \citealt{Thebault:2003}), meaning that objects with large orbital inclinations could be resurfaced more often. 
This brings us back to the decades-long \textit{nature versus nurture} debate about the origins of TNO colors. 
While it has been convincingly demonstrated that distinct color classes of TNOs are associated with distinct orbital distributions \citep{Marsset:2019, AliDib:2021}, meaning that TNO colors are to a large extend dependent on their original formation location, more subtle intraclass spectral variability could very well originate from distinct collisional histories of the class members \citep{DellOro:2013, Abedin:2021}. 


\section{Summary}
\label{sec:summary}

Motivated by earlier studies that found concordance between the color and dynamical distributions of TNOs \citep{Tegler:2000, Peixinho:2008, Tegler:2016,Marsset:2019, AliDib:2021}, we investigated the existence of similar correlations in the new {\it faintIR/brightIR} taxonomic scheme (see companion paper by Fraser et al.). In addition to confirming the distinct orbital distributions of the two composition classes -- the so-called \textit{faintIR} and \textit{brightIR} classes, we further report the detection 
of a correlation between colors and orbital inclinations \textit{within} the \textit{brightIR} class. 
This new color versus dynamical structure in the Kuiper belt remains undetected when using the traditional approach of dividing TNOs into optically \textit{less red} and \textit{very red} objects. 

Considering that: (1) the final excitation state of TNOs correlate with their initial semi-major axis in simulations of Neptune's migration (e.g., \citealt{Kaib:2016, Nesvorny:2016, Nesvorny:2020, AliDib:2021, Pirani:2021}) and (2) intraclass spectral change reflects composition change in the {\it faintIR/brightIR} classification scheme, we propose that the identified color-inclination correlation reflects a composition gradient in the early proto-planetary disk, over the range of heliocentric distances where \textit{brightIR} TNOs originally formed. 
However, additional dynamical work is currently needed to reconcile this scenario with the presence of blue binary contaminants found in the cold classical region \citep{Fraser:2017, Fraser:2021,Nesvorny:2022}, and to differentiate this scenario from the alternative hypothesis of a collisional origin for the intraclass color versus inclination correlation.

\section*{Acknowledgements}

The authors acknowledge the sacred nature of Maunakea, and appreciate the opportunity to observe from the mountain. 
This work is based on observations from the Large and Long Program GN-2014B-LP-1, GN-2015A-LP-1, GN-2015B-LP-1, GN-2016A-LP-1, GN-2016B-LP-1, GN-2017A-LP-1, GN-2018A-Q-118, GN-2018A-Q-223, and GN-2020B-Q-127 obtained at the Gemini Observatory, which is operated by the Association of Universities for Research in Astronomy, Inc., under a cooperative agreement with the NSF on behalf of the Gemini partnership: the National Science Foundation (United States), the National Research Council (Canada), CONICYT (Chile), Ministerio de Ciencia, Tecnolog\'{i}a e Innovaci\'{o}n Productiva (Argentina), and Minist\'{e}rio da Ci\^{e}ncia, Tecnologia e Inova\c{c}\~{a}o (Brazil). 
We thank the staff at Gemini North for their dedicated support of the Col-OSSOS program. 
Data was processed using the Gemini IRAF package. 
STSDAS and PyRAF are products of the Space Telescope Science Institute, which is operated by AURA for NASA.
M.M. was supported by the National Aeronautics and Space Administration under Grant No. 80NSSC18K0849 and 80NSSC18K1004 issued through the Planetary Astronomy Program. MES is supported by the UK Science and Technology Facilities Council (STFC) grant ST/V000691/1. LEB acknowledges funding from the UK STFC Grant Code ST/T506369/1. REP acknowledges funding from NASA Emerging Worlds grant 80NSSC21K0376.

\appendix
\section{Complete dataset}
\label{sec:appA}

The physical, orbital and spectral properties, and surface classification of TNOs from the Col-OSSOS and H/WTSOSS datasets are provided in Table\,\ref{tab:summary}. Similar data for the MBOSS dataset are provided in Table\,\ref{tab:summary_mboss}.

\startlongtable
\begin{deluxetable}{lccccccccccccccc}
\tabletypesize{ \footnotesize}
\tablecaption{Physical, orbital and spectral properties of TNOs from Col-OSSOS and H/WTSOSS.}
\tablehead{Object &    OSSOS &     $H_{mag}$ &        $a$ &        $e$ &        $i$ &       $i_{free}^{\dagger}$ &      visible spt. slope &       NIR spt. slope &   \multicolumn{2}{c}{Surface Class$^{\dagger\dagger}$} &   Ref.$^{\ddagger}$ \\
 &    id &      &        [au] &    &       [degree] &       [degree] &       [\%/($10^{-3}\,$\AA)] &       [\%/($10^{-3}\,$\AA)] &   Bir/Fir & LR/VR &
}
\startdata
       1993 HA2 &           - &     9.60 &    24.70 &     0.52 &    15.60 &        - &    43.10 $\pm$     3.10 &     8.90 $\pm$     1.60 &      Fir &         VR &          1 \\
       1995 DW2 &           - &     8.60 &    25.10 &     0.24 &     4.14 &        - &     5.50 $\pm$     2.60 &    10.00 $\pm$     3.20 &      Bir &         LR &          1 \\
       1995 QY9 &     o4h69PD &     7.68 &    39.64 &     0.26 &     4.83 &        - &    18.20 $\pm$     1.30 &     9.40 $\pm$     1.50 &      Bir &         LR &          2 \\
     1998 SM165 &           - &     5.81 &    48.00 &     0.38 &    13.50 &        - &    24.60 $\pm$     2.20 &    13.60 $\pm$     1.30 &      Bir &         VR &          1 \\
      1998 TF35 &           - &     8.80 &    26.10 &     0.38 &    12.60 &        - &    35.00 $\pm$     1.30 &     9.60 $\pm$     1.20 &      Fir &         VR &          1 \\
      1998 WW24 &           - &     8.30 &    39.40 &     0.23 &    13.90 &        - &    18.00 $\pm$     4.60 &     8.60 $\pm$     5.30 &      Bir &         LR &          1 \\
      1998 WY24 &           - &     6.82 &    43.00 &     0.04 &     1.91 &     0.76 &    19.40 $\pm$     3.60 &     7.90 $\pm$     2.80 &      Fir &         LR &          1 \\
     1999 CL119 &           - &     6.12 &    46.60 &     0.00 &    23.40 &    21.82 &    16.60 $\pm$     1.90 &     1.40 $\pm$     1.20 &      Bir &         LR &          1 \\
\enddata
\tablenotetext{}{$^*$Object with inconsistent surface classification between the different visits; see discussion by Fraser et al. in companion paper.}
\tablenotetext{}{${}^{\dagger}$Free inclination values from \citet{Huang:2022}.}
\tablenotetext{}{${}^{\dagger\dagger}$Surface classification: Bir=BrightIR, Fir=FaintIR, LR=Less Red, VR=Very Red. The Bir/Fir class assignment was performed based on PC values as described in Section\,\ref{sec:taxonomy}. The LR/VR class assignment was performed based on the optical slope values, with the limit between LR and VR objects placed at 20.6\%/$10^3\,\rm{\AA}$ following \citet{Marsset:2019}.}
\tablenotetext{}{${}^{\ddagger}$Origin of spectral slope measurement: 1=H/WTSOSS, 2=Col-OSSOS.}
\label{tab:summary}
\end{deluxetable}

\startlongtable
\begin{deluxetable}{lccccccccccccc}
\tabletypesize{ \footnotesize}
\tablecaption{Orbital and photometric properties of TNOs from MBOSS.}
\tablehead{Object &     $H_{mag}$ &        $a$ &        $e$ &        $i$ &       $i_{free}^{\dagger}$ &       visible spt. slope &       NIR spt. slope &   \multicolumn{2}{c}{Surface Class$^{\dagger\dagger}$} \\
  &      &        [au] &    &       [degree] &       [degree]  &       [\%/($10^{-3}\,$\AA)] &       [\%/($10^{-3}\,$\AA)] &   Bir/Fir & LR/VR
}
\startdata
      1977 UB &     6.09 &    13.70 &     0.38 &     6.92 &        - &     3.30 $\pm$     1.00 &    -0.50 $\pm$     1.60 &      Bir &       LR \\
      1992 AD &     7.14 &    20.30 &     0.57 &    24.60 &        - &    47.60 $\pm$     9.90 &    38.30 $\pm$     7.30 &      Fir &       VR \\
     1993 HA2 &     9.05 &    24.70 &     0.52 &    15.60 &        - &    47.60 $\pm$     9.90 &    19.40 $\pm$     3.80 &      Fir &       VR \\
      1993 SC &     6.75 &    39.70 &     0.19 &     5.15 &     6.70 &    37.00 $\pm$    10.60 &    20.60 $\pm$    11.80 &      Fir &       VR \\
      1994 TB &     7.51 &    39.60 &     0.32 &    12.10 &    14.23 &    34.40 $\pm$    16.00 &    28.50 $\pm$    14.60 &      Fir &       VR \\
     1995 DW2 &     9.24 &    25.10 &     0.24 &     4.14 &        - &    10.40 $\pm$    15.50 &     0.90 $\pm$     3.30 &      Bir &       LR \\
      1995 GO &     9.00 &    18.00 &     0.62 &    17.60 &        - &    13.50 $\pm$     3.30 &     7.50 $\pm$     1.50 &      Bir &       LR \\
     1995 QY9 &     7.37 &    39.90 &     0.27 &     4.83 &        - &    15.70 $\pm$    10.20 &    19.30 $\pm$    11.60 &      Bir &       LR \\
\enddata
\tablenotetext{}{${}^{\dagger}$Free inclination values from \citet{Huang:2022}.}
\tablenotetext{}{${}^{\dagger\dagger}$Surface classification: Bir=BrightIR, Fir=FaintIR, LR=Less Red, VR=Very Red. The Bir/Fir class assignment was performed based on PC values as described in Section\,\ref{sec:taxonomy}. The LR/VR class assignment was performed based on the optical slope values, with the limit between LR and VR objects placed at 20.6\%/$10^3\,\rm{\AA}$ following \citet{Marsset:2019}.}
\label{tab:summary_mboss}
\end{deluxetable}

\section{Results of statistical tests}
\label{sec:appB}

The results of the Pearson and Spearman $r$ correlation tests between the orbital properties ($a$, $e$, $i$, $i_{free}$) and spectral properties (optical and near-infrared spectral slopes) of TNOs are provided for the Col-OSSOS+H/WTSOSS dataset, the MBOSS dataset, and the Col-OSSOS+H/WTSOSS in Table\,\ref{tab:correlations1}, Table\,\ref{tab:correlations2} and Table\,\ref{tab:correlations3}, respectively.

\startlongtable
\begin{deluxetable}{l | cc | cc | cc | cc}
\tabletypesize{ \small}
\tablecaption{Summary of Pearson/Spearman probabilities of correlations (in \%) for the Col-OSSOS+H/WTSOSS dataset.}
\tablehead{ & \multicolumn{2}{c |}{Visible Spt. Slope} & \multicolumn{2}{c |}{NIR Spt. Slope} & \multicolumn{2}{c |}{PC1} & \multicolumn{2}{c}{PC2} \\
  &  BrightIR & FaintIR &  BrightIR & FaintIR &  BrightIR & FaintIR &  BrightIR & FaintIR }
\startdata
\multicolumn{1}{c }{} & \multicolumn{8}{c}{Complete dataset (N=189)}\\
\hline
Semi-major axis  & 61.9/96.5 & 62.9/7.4 & 11.5/19.0 & 63.1/0.9 & 29.1/38.7 & 65.1/9.5 & 68.7/96.3 & 7.6/3.8\\
Inclination      & 99.5/99.5 & 98.4/92.8 & 94.1/81.9 & 71.7/73.6 & 96.1/90.7 & 75.0/84.9 & 75.5/77.6 & 13.9/25.8\\
Free inclination & 99.3/99.2 & 99.0/87.3 & 92.5/78.5 & 73.5/28.0 & 95.2/89.7 & 78.3/56.8 & 77.3/82.4 & 23.5/14.9\\
Eccentricity     & 20.8/46.2 & 28.2/68.7 & 42.9/37.0 & 87.3/40.1 & 25.3/52.8 & 78.4/1.4 & 26.5/54.6 & 78.0/90.8\\
\hline
\multicolumn{1}{c }{} & \multicolumn{8}{c}{Excluding Large Objects ($H_{mag}<$5; N=182)}\\
\hline
Semi-major axis  & 42.4/92.8 & 64.0/25.8 & 2.8/1.6 & 63.1/8.9 & 35.0/55.2 & 65.3/18.1 & 43.8/91.5 & 8.8/19.4\\
Inclination      & 99.6/99.7 & 98.8/95.3 & 92.1/76.1 & 72.3/74.9 & 95.0/87.3 & 76.3/86.2 & 85.2/88.9 & 18.7/17.2\\
\enddata
\label{tab:correlations1}
\end{deluxetable}

\startlongtable
\begin{deluxetable}{l | cc | cc | cc | cc}
\tabletypesize{ \small}
\tablecaption{Same as Table~\ref{tab:correlations1} for the MBOSS dataset.}
\tablehead{ & \multicolumn{2}{c |}{Visible Spt. Slope} & \multicolumn{2}{c |}{NIR Spt. Slope} & \multicolumn{2}{c |}{PC1} & \multicolumn{2}{c}{PC2} \\
  &  BrightIR & FaintIR &  BrightIR & FaintIR &  BrightIR & FaintIR &  BrightIR & FaintIR }
\startdata
\multicolumn{1}{c }{} & \multicolumn{8}{c}{Complete dataset (N=55)}\\
\hline
Semi-major axis  & 4.4/22.6 & 62.3/7.8 & 26.9/28.1 & 1.8/64.9 & 13.5/30.5 & 12.1/64.9 & 6.4/22.4 & 74.3/42.0\\
Inclination      & 99.3/99.1 & 54.8/42.1 & 94.8/95.5 & 80.5/83.6 & 98.6/96.3 & 84.4/83.6 & 98.0/98.3 & 9.3/2.0\\
Free inclination & 99.1/99.1 & 53.4/43.8 & 94.7/95.7 & 81.4/84.0 & 98.6/96.5 & 85.6/84.0 & 97.3/98.0 & 5.7/3.0\\
Eccentricity     & 17.9/12.5 & 42.7/60.8 & 66.5/53.7 & 18.4/89.1 & 42.8/47.9 & 45.3/89.1 & 55.5/10.3 & 32.3/45.5\\
\hline
\multicolumn{1}{c }{} & \multicolumn{8}{c}{Excluding Large Objects ($H_{mag}<$5; N=35)}\\
\hline
Semi-major axis  & 42.5/78.2 & 89.5/80.1 & 72.6/96.5 & 44.3/18.8 & 79.5/96.4 & 44.3/22.1 & 14.4/64.1 & 92.2/93.0\\
Inclination      & 61.2/80.2 & 24.8/0.0 & 10.5/16.5 & 26.9/1.7 & 21.6/11.6 & 27.0/6.9 & 71.9/80.5 & 28.7/21.3\\
\enddata
\label{tab:correlations2}
\end{deluxetable}

\startlongtable
\begin{deluxetable}{l | cc | cc | cc | cc}
\tabletypesize{ \small}
\tablecaption{Same as Table~\ref{tab:correlations1} for combined Col-OSSOS+H/WTSOSS and MBOSS datasets.}
\tablehead{ & \multicolumn{2}{c |}{Visible Spt. Slope} & \multicolumn{2}{c |}{NIR Spt. Slope} & \multicolumn{2}{c |}{PC1} & \multicolumn{2}{c}{PC2} \\
  &  BrightIR & FaintIR &  BrightIR & FaintIR &  BrightIR & FaintIR &  BrightIR & FaintIR }
\startdata
\multicolumn{1}{c }{} & \multicolumn{8}{c}{Complete dataset (N=235)}\\
\hline
Semi-major axis  & 38.7/94.4 & 65.9/48.2 & 21.2/52.7 & 63.9/87.5 & 77.1/87.7 & 58.8/56.4 & 25.2/73.8 & 3.6/10.7\\
Inclination      & 100.0/100.0 & 98.6/97.3 & 99.4/95.9 & 99.0/97.5 & 96.6/96.7 & 71.9/78.5 & 92.9/91.1 & 10.8/38.4\\
Free inclination & 100.0/100.0 & 99.2/96.9 & 99.3/95.2 & 99.1/89.0 & 95.0/95.2 & 73.8/45.0 & 94.3/93.9 & 1.2/14.9\\
Eccentricity     & 77.2/77.7 & 91.6/96.9 & 60.8/64.3 & 94.1/93.0 & 5.3/30.6 & 92.5/68.9 & 38.0/20.4 & 73.2/88.1\\
\hline
\multicolumn{1}{c }{} & \multicolumn{8}{c}{Excluding Large Objects ($H_{mag}<$5; N=210)}\\
\hline
Semi-major axis  & 27.0/83.3 & 71.0/56.9 & 43.5/71.6 & 69.0/94.6 & 92.1/97.0 & 60.0/69.6 & 16.5/19.2 & 1.0/21.8\\
Inclination      & 99.9/100.0 & 99.9/99.2 & 92.0/66.5 & 98.8/97.5 & 74.6/76.6 & 70.3/77.9 & 96.3/97.3 & 23.2/2.8\\
\enddata
\label{tab:correlations3}
\end{deluxetable}

\bibliography{references} 
\bibliographystyle{aasjournal}

\end{document}